\documentclass[prl, 10pt,aps,nofootinbib, floatfix, notitlepage,twocolumn,longbibliography]{revtex4-2}
\usepackage{amsmath,mathrsfs,amsfonts,amssymb,bm}
\usepackage{dcolumn,graphicx,xcolor,mathtools,booktabs,microtype}
\usepackage[utf8]{inputenc}
\definecolor{navyblue}{rgb}{0.0, 0.0, 0.5}
\usepackage[colorlinks=true, linkcolor=navyblue, citecolor=navyblue, urlcolor=navyblue]{hyperref}
\usepackage{orcidlink}
\newcommand{\trm}[1]{{\mbox{\tiny\rm{#1}}}}

\newcommand{\bbl}{\mathbf{B\!-\!L}} 
\newcommand{\itbl}{\textit{B\!\,-\!\,L}}
\newcommand{\rmbl}{\mathrm{B\!-\!L}}

\def\be{\begin{equation}}
\def\ee{\end{equation}}
\def\bea{\begin{align}}
\def\eea{\end{align}}
\def\bea{\begin{eqnarray}}
\def\eea{\end{eqnarray}}
\def\bi{\begin{itemize}}
\def\ei{\end{itemize}}
\begin{document}
\title{Axion-Sourced Gravitational Waves in $\bbl$ Hybrid Inflation}

\author{Ahmed Ayad\,\orcidlink{0000-0001-8774-1394}}
\author{Shaaban Khalil\,\orcidlink{0000-0003-1950-4674}}
\affiliation{Center for Fundamental Physics, Zewail City of Science and Technology, 6th October City, Giza 12588, Egypt} 
\date{\today}
\begin{abstract}
Minimal supersymmetric $\rmbl$ hybrid inflation predicts a negligible vacuum tensor-to-scalar ratio $r_{\rm vac}\sim 10^{-8}$ due to its extremely flat potential. In this letter, we show that adding a spectator pseudoscalar axion-like field coupled to the $\rm U(1)_\trm{B-L}$ gauge sector via a Chern-Simons term circumvents this suppression. The rolling axion triggers a tachyonic instability for one gauge field helicity, exponentially amplifying gauge fluctuations. These sourced modes generate a stochastic gravitational-wave background with tensor power spectrum $\mathcal{P}_{T}^{\rm src}\propto e^{4\pi\xi}/\xi^{6}$, where $\xi = \alpha \dot{\phi} / (2 f_a H)$ is the gauge-field instability parameter. For $\xi\sim3.3$--$3.6$, the tensor-to-scalar ratio reaches $r\sim10^{-3}$, within the sensitivity of LiteBIRD. For $\xi\sim3.6$, the gravitational-wave spectrum develops a peaked shape that enters the LISA sensitivity band (peaking at frequencies around $10^{-3}$ Hz) with an amplitude $h^2\Omega_{\mathrm{GW}}\sim2.5\times10^{-13}$. The mechanism predicts chiral gravitational waves, a broken consistency relation $r=16\epsilon$, and a distinctive $r$--$\Omega_{\rm GW}$ correlation that could be tested by future CMB and interferometer observations.
\end{abstract}
\maketitle
\textit{Introduction---}Supersymmetric hybrid inflation provides a well-motivated framework for embedding cosmic inflation within particle-physics models, particularly those based on a gauged $\rm U(1)_\trm{B-L}$ symmetry~\cite{Dvali:1994ms, Lazarides:1995vr}. In its minimal realization, however, the predicted primordial tensor signal is extremely suppressed~\cite{Rehman:2009nq, Wu:2016fzp}. The flatness of the inflationary potential leads to a tiny slow-roll parameter $\epsilon$ at CMB horizon exit, and therefore to a negligible tensor-to-scalar ratio, $r \simeq 16\epsilon$~\cite{Mukhanov:2005sc, Liddle:2000cg, Linde:1990flp}. Consequently, the standard vacuum contribution to gravitational waves (GW) in $\rmbl$ hybrid inflation is effectively unobservable.

In this letter, we show that this conclusion can be changed qualitatively without modifying the core structure of the model. We supplement the inflationary sector with a spectator pseudoscalar axion-like field coupled to the $\rm U(1)_\trm{B-L}$ gauge field through a Chern--Simons interaction~\cite{Barnaby:2011qe}. A spectator axion coupled to an $\rm SU(2)$ gauge field yields a similar effect in~\cite{Barnaby:2011qe}. Traditionally, axions were first introduced to resolve the strong CP problem~\cite{Peccei:1977hh, Peccei:1977ur, Weinberg:1977ma, Wilczek:1977pj}, and axion‑like particles appear generically in string compactifications and other SM extensions~\cite{Anselm:1981aw,Kim:1986ax,Jaeckel:2010ni}. A time-dependent axion background induces a tachyonic instability in one gauge field helicity, exponentially amplifying gauge fluctuations~\cite{Anber:2006xt,Barnaby:2011qe}. These amplified modes generate anisotropic stress that sources a stochastic background of GW, see e.g. Ref.~\cite{Badger:2024ekb}.

A key feature of this mechanism is that it relies only on the existing $\rmbl$ gauge sector, without introducing an additional hidden sector. The inflationary background is driven by the singlet field $S$, while tensor production originates from the axion--gauge sector and is largely independent of slow-roll dynamics. The sourced tensor signal can greatly exceed the vacuum contribution and exhibits distinctive signatures. Since only one gauge field helicity is amplified, the resulting GW are chiral and can generate parity-violating signatures~\cite{Barnaby:2011qe, Dimastrogiovanni:2016fuu}. The tensor spectrum may also deviate from scale invariance, developing a localized peak depending on the axion evolution. The tensor amplitude is controlled by the instability parameter $\xi = \alpha \dot{\phi}/(2f_a H)$ (with $\alpha$ a dimensionless coupling, $f_a$ the axion decay constant, and $H$ the Hubble parameter), not by the slow-roll parameter $\epsilon$. Therefore, the standard consistency relation $r = 16\epsilon$ no longer applies~\cite{Dimastrogiovanni:2016fuu}.

This framework opens multiple observational windows for $\rmbl$ hybrid inflation. Large-scale tensor modes may be probed by future CMB experiments such as LiteBIRD~\cite{LiteBIRD:2022cnt, LiteBIRD:2023iei}, while the same mechanism generates a peaked stochastic GW background in the LISA band~\cite{Caprini:2019pxz, Boileau:2020rpg}. Motivated by these observational prospects, we quantify this mechanism within the $\rmbl$ hybrid inflation framework by deriving the coupled axion--gauge dynamics, computing the sourced tensor spectrum, and identifying parameter regions compatible with current constraints and observable GW signals.
\vskip 0.1cm
\textit{Vacuum tensor modes in $\itbl$ inflation---}A minimal supersymmetric $\rmbl$ hybrid inflation model is described by the superpotential~\cite{Dvali:1994ms, Lazarides:1995vr}
\begin{equation}
W_{\rm inf} = \kappa S(\Phi\bar\Phi-v_\trm{B-L}^2)\,,
\end{equation}
where $S$ is the gauge-singlet inflaton superfield, $\Phi$ and $\bar\Phi$ are oppositely charged waterfall fields, $\kappa$ is a dimensionless coupling, and $v_\trm{B-L}$ is the $\rm U(1)_\trm{B-L}$ breaking VEV. For related studies of GW from preheating and cosmic strings in $\rmbl$ hybrid inflation, see e.g. Refs.~\cite{Buchmuller:2012wn, Buchmuller:2013lra}. The bosonic sector contains the scalar fields and the $\rm U(1)_\trm{B-L}$ gauge field $Z'_\mu$,
\begin{align}
\mathcal{L}_{\rm BL\,hybrid} &= |\partial_\mu S|^2 + |D_\mu\Phi|^2 + |D_\mu\bar\Phi|^2 -\frac14 Z'_{\mu\nu}Z'^{\mu\nu} \nonumber\\
&\quad - V_{\rm BL\,hybrid}\,,
\end{align}
where $D_\mu$ is the covariant derivative and $Z'_{\mu\nu}$ the $\rm U(1)_\trm{B-L}$ field strength tensor. The scalar potential is
\begin{align}
V_{\rm BL\,hybrid} &= \kappa^2 |\Phi\bar\Phi-v_\trm{B-L}^2|^2 + \kappa^2 |S|^2(|\Phi|^2+|\bar\Phi|^2) \nonumber\\
&\quad + \frac{g_\trm{B-L}^2}{2} (|\Phi|^2-|\bar\Phi|^2)^2 \,.
\end{align}
Along the inflationary trajectory, $\Phi = \bar{\Phi} = 0$, the tree‑level potential is approximately flat, $V_{\rm inf}\simeq \kappa^2 v_\trm{B-L}^4$, and the slow-roll dynamics is generated by radiative Coleman--Weinberg corrections~\cite{Coleman:1973jx, Dvali:1994ms},
\begin{equation}
V_{\rm eff}(S) \simeq \kappa^2 v_\trm{B-L}^4 + \frac{\kappa^4 v_\trm{B-L}^4}{8\pi^2} \ln\!\left(\frac{|S|}{Q}\right)\,.
\end{equation}
where $Q$ is the renormalization scale. Inflation ends when the inflaton reaches the critical value $|S|\sim v_\trm{B-L}$, at which point the waterfall fields become tachyonic and $\rm U(1)_\trm{B-L}$ is spontaneously broken.

Tensor perturbations are defined through
\begin{equation}
ds^2 = -dt^2 + a^2(t)(\delta_{ij}+h_{ij})dx^i dx^j\,,
\end{equation}
where $a(t)$ is the scale factor, $\delta_{ij}$ is the Kronecker delta, and $h_{ij}$ is transverse and traceless. Expanding the Einstein--Hilbert action to quadratic order yields~\cite{Mukhanov:2005sc}
\begin{equation}
S_{\rm T}^{(2)} = \frac{M_{\rm Pl}^2}{8} \int d\tau\, d^3x\, a^2 \Big[(h_{ij}')^2-(\nabla h_{ij})^2\Big]\,,
\end{equation}
where $M_{\rm Pl}$ the reduced Planck mass, $\tau$ is conformal time and primes denote $d/d\tau$. After canonical normalization, the vacuum tensor power spectrum is
\begin{equation}
\mathcal{P}_{\rm T}^{\rm vac} \simeq \frac{2H_{\ast}^2}{\pi^2 M_{\rm Pl}^2}\,,
\end{equation}
where asterisk $\ast$ denotes the value evaluated at horizon crossing of the observed CMB scale~\cite{Starobinsky:1979ty}. The scalar curvature perturbation is
\begin{equation}
\mathcal{P}_\zeta \simeq \frac{H_{\ast}^2} {8\pi^2 \epsilon_{\ast}M_{\rm Pl}^2}\,,
\end{equation}
leading to the standard consistency relation $r_{\rm vac} = 16\epsilon_{\ast}$.
Because the inflationary potential is extremely flat in $\rmbl$ hybrid inflation, the slow-roll parameter $\epsilon_{\ast}$ is typically tiny, implying $r_{\rm vac}\ll 1$. Therefore, the vacuum contribution to primordial GW is negligible, motivating the introduction of additional non-vacuum sources of tensor perturbations, such as the axion--gauge mechanism discussed below.
\vskip 0.1cm
\textit{Axion--gauge instability in $\itbl$ inflation---}To generate a sizable tensor signal beyond the suppressed vacuum contribution of hybrid inflation, we introduce a spectator pseudoscalar axion-like field $\phi$ coupled to the $\rm U(1)_\trm{B-L}$ gauge field through a Chern–Simons interaction~\cite{Anber:2006xt, Barnaby:2011qe}. During inflation, we treat $\phi$ as a classical homogeneous background field, $\phi = \phi(t)$, neglecting its spatial fluctuations for the purpose of gauge field production. Their backreaction on scalar perturbations will be discussed later. The relevant effective Lagrangian is
\begin{equation}
\mathcal{L} = \mathcal{L}_{\rm BL\,hybrid} + \frac{1}{2}\partial_\mu \phi \,\partial^\mu \phi - V(\phi) - \frac{\alpha}{4f_a}\, \phi \,Z'_{\mu\nu}\tilde Z'^{\mu\nu}\,.
\label{eq:lagrangian_axion}
\end{equation}
The axion does not couple directly to the waterfall fields $\Phi,\bar{\Phi}$; therefore it does not affect the $\rm U(1)_\trm{B-L}$ symmetry breaking scale $v_\trm{B-L}$. The axion potential is taken to be the standard cosine form~\cite{Freese:1990rb},
\begin{equation}
V(\phi) = \Lambda^4 \left(1 - \cos \frac{\phi}{f_a}\right)\,,
\label{eq:axion_potential}
\end{equation}
where $\Lambda$ is some non-perturbatively generated scale. During inflation we assume $V(\phi)\ll V_{\rm inf}$ so that the axion is a true spectator, consistent with the energy‑density condition $\rho_{\phi} \ll V_{\rm inf}$. The axion need not slow‑roll for many e‑folds; a few e‑folds are sufficient to source observable GW on CMB scales. From this Lagrangian, the homogeneous background axion obeys the equation of motion obtained from varying the action of Lagrangian~\eqref{eq:lagrangian_axion},
\begin{equation}
\frac{1}{\sqrt{-g}}\partial_\mu\Bigl(\sqrt{-g}\,g^{\mu\nu}\partial_\nu \phi\Bigr) + V'(\phi) = \frac{\alpha}{4f_a}Z'_{\mu\nu}\tilde Z'^{\mu\nu}\,.
\label{eq:axioneom}
\end{equation}
In a flat FRW background this reduces to
\begin{equation}
\ddot{\phi} + 3H\dot{\phi} + V' = \frac{\alpha}{f_a}\bigl\langle \mathbf{E}_{\trm{B-L}}\cdot\mathbf{B}_{\trm{B-L}}\bigr\rangle\,,
\label{eq:axionfrw}
\end{equation}
where we have used $Z'_{\mu\nu}\tilde Z'^{\mu\nu} = -4\,\mathbf{E}_{\trm{B-L}}\cdot\mathbf{B}_{\trm{B-L}}$, and the expectation value denotes the averaged backreaction of the produced gauge quanta on the axion. 

The gauge field equation including the Chern–Simons term follows from varying the action,
\begin{equation}
\nabla_\mu Z'^{\mu\nu} + \frac{\alpha}{f_a}(\partial_\mu \phi)\tilde Z'^{\mu\nu} = 0\,,
\label{eq:gaugeeom}
\end{equation}
where matter currents have been neglected during inflation. Working in conformal time $\tau$ and Coulomb gauge, $Z_{0}^{\prime} = 0,\; \nabla \cdot \mathbf{Z}^{\prime} = 0$, the equation becomes
\begin{equation}
\mathbf{Z}^{\prime\prime} - \nabla^2\mathbf{Z}' + \frac{\alpha \phi'}{f_a}\nabla\times\mathbf{Z}' = 0\,.
\label{eq:wave}
\end{equation}
Decomposing into circular polarization modes with comoving wavenumber $k = |\mathbf{k}|$,
\begin{equation}
Z_\lambda^{\prime\prime} + \left(k^2 + \lambda k\,\frac{\alpha \phi'}{f_a}\right)Z_\lambda = 0\,,
\label{eq:mode}
\end{equation}
where $\lambda = \pm$ labels the helicity. We introduce the instability parameter~\cite{Barnaby:2011qe, Anber:2006xt}
\begin{equation}
\xi \equiv \frac{\alpha\dot{\phi}}{2f_{a}H}\,,
\label{eq:xi}
\end{equation}
that quantifies how efficiently the gauge field is produced. With the approximation $\langle\mathbf{E}\cdot\mathbf{B}\rangle \approx \frac{H^4}{8\pi^2}\frac{e^{2\pi\xi}}{\xi}$~\cite{Anber:2006xt} and $\dot\phi = 2f_a H\xi/\alpha$, the backreaction contribution in Eq.~\eqref{eq:axionfrw} remains subdominant compared to $3H\dot\phi$ for $\xi\lesssim4.3$; for our benchmark $H/M_{\rm Pl}\sim10^{-5}$ and $f_a\sim10^{16}$~GeV, the ratio $R\equiv\frac{\alpha}{f_a}\langle\mathbf{E}\cdot\mathbf{B}\rangle/(3H\dot\phi)\sim0.3$ at $\xi=3.6$. Over the few e-folds relevant for CMB scales, $\xi$ evolves slowly ($|\dot\xi/(H\xi)|\ll1$), justifying the constant-$\xi$ approximation at CMB scales. However, to capture the frequency-dependent GW signal at interferometer scales, we must account for the scale dependence of $\xi$, which we model in the next section.

During quasi‑de Sitter expansion, $a(\tau) \simeq -1/(H\tau)$. Using $\phi' = a\dot{\phi} \simeq -\dot{\phi}/(H\tau)$, the combination that appears in the mode equation becomes
\begin{equation}
\frac{\alpha \phi'}{f_a} = -\frac{\alpha\dot{\phi}}{f_a H \tau} = -\frac{2\xi}{\tau}\,.
\end{equation}
Hence Eq.~\eqref{eq:mode} can be written as
\begin{equation}
Z_{\pm}^{\prime\prime} + \left(k^{2} \pm 2k\,\frac{\xi}{\tau}\right)Z_{\pm} = 0\,.
\label{eq:modes}
\end{equation}
For a slowly rolling axion, $\xi$ remains approximately constant during the few e‑folds around horizon crossing, justifying its treatment as a constant when computing the power spectrum. For $\dot{\phi}>0$ (hence $\xi>0$), the mode with $\lambda=+$ experiences the tachyonic instability. For $\xi \gtrsim \mathcal{O}(1)$, the effective frequency squared for the $+$ helicity becomes negative for modes with $k|\tau| \lesssim 2\xi$, triggering a tachyonic instability and exponential amplification, $Z_{+}\propto e^{\pi \xi}$, while the opposite helicity remains close to its vacuum amplitude~\cite{Anber:2006xt}. The resulting gauge‑field power spectrum scales as $\langle Z_+ Z_+\rangle \propto e^{2\pi\xi}$, producing a highly helical state. Thus the rolling axion acts as a catalyst. It amplifies one helicity of the $\rmbl$ gauge field, which in turn generates anisotropic stress. That stress sources GW, producing a stochastic background that is chiral, potentially non‑Gaussian, and far larger than the negligible vacuum signal of pure hybrid inflation. We now compute this sourced contribution quantitatively.
\vskip 0.1cm
\textit{Sourced tensor modes from axion--$\itbl$ coupling---}Tensor perturbations $h_{ij}$ are sourced by the transverse‑traceless (TT) part of the anisotropic stress tensor~\cite{Mukhanov:2005sc},
\begin{equation}
\ddot h_{ij} + 3H\dot h_{ij} - \frac{\nabla^2}{a^2}h_{ij} = \frac{2}{M_{\rm Pl}^2}\,\Pi_{ij}^{\rm TT}\,.
\label{eq:tensoreom}
\end{equation}
In this setup, the dominant contribution arises from the amplified $\rm U(1)_\trm{B-L}$ gauge field, whose anisotropic stress tensor $\Pi_{ij}$ is given by
\begin{equation}
\Pi_{ij}\simeq E_i^{\trm{B-L}}E_j^{\trm{B-L}} + B_i^{\trm{B-L}}B_j^{\trm{B-L}}
- \frac{1}{3}\delta_{ij}\bigl(\mathbf{E}_\trm{B-L}^2 + \mathbf{B}_\trm{B-L}^2\bigr)\,.
\label{eq:anisotropic}
\end{equation}
The source originates from the tachyonic amplification of one helicity of the gauge field induced by the rolling spectator axion via the Chern–Simons interaction~\cite{Anber:2006xt, Barnaby:2011qe}. As a result, the tensor power spectrum receives both vacuum and sourced contributions,
\begin{equation}
\mathcal{P}_{\rm T} = \mathcal{P}_{\rm T}^{\rm vac} + \mathcal{P}_{\rm T}^{\rm src}\,,
\label{eq:totalpower}
\end{equation}
with the sourced component given by~\cite{Domcke:2016bkh}
\begin{equation}
\mathcal{P}_{\rm T}^{\rm src} = C_{\rm T} \left(\frac{H}{M_{\rm Pl}}\right)^4 \frac{e^{4\pi\xi}}{\pi^2 \xi^6}\,,
\label{eq:srcpower}
\end{equation}
where the instability parameter $\xi$ is defined in Eq.~\eqref{eq:xi}. The exponential factor $e^{4\pi\xi}$ follows from the tachyonic growth $Z_+ \propto e^{\pi\xi}$ and the quadratic dependence of the anisotropic stress on the $\rm U(1)_\trm{B-L}$ gauge field. The constant $C_{\rm T}$ arises from the momentum‑space integration and is typically small; the exponential factor nonetheless prevails for $\xi \gtrsim \mathcal{O}(1)$, enabling a sizable tensor signal even when the slow-roll parameter $\epsilon$ is extremely small.

The sourced tensor signal can be expressed in terms of the present‑day stochastic GW spectrum as a function of comoving wavenumber $k$ using
\begin{equation}
h^2\Omega_{\trm{GW}}(k) \simeq \frac{h^2\Omega_{\rm r}}{12}\,\mathcal{P}_{\rm T}(k)\,,
\label{eq:omegagw_k}
\end{equation}
where $\Omega_{\rm r}$ is the present radiation density parameter and $h$ here denotes the dimensionless Hubble parameter. For modes re‑entering during radiation domination, the relation holds for $k$ corresponding to frequencies above $\sim 10^{-15}$ Hz. To connect to observations at different frequencies, we convert comoving wavenumber $k$ to present frequency $f$,
\begin{equation}
f = \frac{k}{2\pi a_0} \approx 1.55\times10^{-15}\,{\rm Hz}\;\left(\frac{k}{{\rm Mpc}^{-1}}\right)\,,
\label{eq:frequency}
\end{equation}
where $a_0$ is the present scale factor. The CMB scale $k_*=0.05\,{\rm Mpc}^{-1}$ corresponds to $f_*\simeq 7.75 \times10^{-17}$ Hz, while LISA is sensitive around $10^{-3}$ Hz, i.e., $k\sim 10^{15}\,{\rm Mpc}^{-1}$, far inside the horizon at the end of inflation. The amplitude at these smaller scales depends on the evolution of $\xi$. For a slowly rolling axion, $\xi$ typically grows as we go to smaller scales (earlier times)~\cite{Barnaby:2011qe}, which gives a blue tilt to the sourced tensor spectrum. In the strong backreaction regime, however, $\xi(f)$ develops a log-normal peak, $\xi(f)=\xi_{\rm peak}\exp[-(\ln(f/f_{\rm peak}))^2/(2\sigma^2)]$, with $f_{\rm peak}=10^{-3}$~Hz. Parametrising $\mathcal{P}_T^{\rm src}(k) \propto k^{n_T^{\rm src}}$ with $n_T^{\rm src}>0$, the enhancement factor from CMB to LISA scales is $(k_{\rm LISA}/k_*)^{n_T^{\rm src}}$. For a representative blue tilt $n_T^{\rm src}\simeq 1$, this factor is $\sim (10^{15}/0.05)^{1} \approx 2\times10^{16}$. Such an enhancement easily compensates the small coefficient $C_T$ and brings $h^2\Omega_{\rm GW}$ into the $10^{-13}$–$10^{-12}$ range for $\xi\simeq3.3$–$3.6$ at LISA frequencies ($f\sim 10^{-3}$ Hz)~\cite{Domcke:2016bkh}, with the spectrum exponentially suppressed above $f\sim 10^{-3}$ Hz as the gauge field production shuts off. A full numerical integration would refine the prediction, but the qualitative conclusion is that the same mechanism can be tested by both LiteBIRD and LISA~\cite{Badger:2024ekb}.

Since the axion is subdominant, $\rho_{\phi} \ll V_{\rm inf}$, inflation is still driven by the hybrid sector. The axion acts only as a driver for gauge‑field amplification, while the gauge field itself provides the anisotropic stress that sources GW. This allows for a sizable tensor signal even when the slow‑roll parameter $\epsilon$ is very small. For illustration, we take $H/M_{\rm Pl}=1.4\times10^{-5}$ and the normalisation $C_T = 8.6\times10^{-7}$ from Ref.~\cite{Domcke:2016bkh}, with $r_{\rm vac}=10^{-8}$ and $\mathcal{P}_\zeta=2.1\times10^{-9}$~\cite{Planck:2018jri}. These benchmark values illustrate the parametric dependence on $\xi$ for concreteness. The resulting behaviour is illustrated in the figures below.

\begin{figure}[th!]
\centering
\includegraphics[width=\columnwidth]{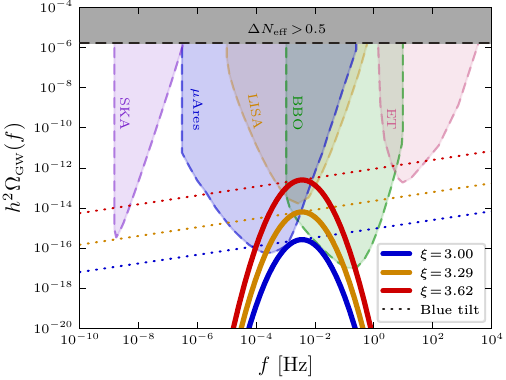}
\caption{Gravitational-wave spectrum $h^2\Omega_{\rm GW}$ as a function of frequency $f$ for $\xi=3.0$, $3.29$, and $3.62$ (solid: includes a Gaussian peak; dotted: only blue‑tilted). The spectrum peaks at $f\sim10^{-3}$ Hz, entering the LISA band for $\xi\gtrsim3.3$. Horizontal dashed line: BBN bound; shaded: LISA, BBO, ET, $\mu$Ares, SKA.}
\label{fig:parameter_space}
\end{figure}

Figure~\ref{fig:r_vs_xi} shows that the tensor‑to‑scalar ratio grows rapidly with $\xi$, transitioning from the negligible vacuum value to potentially observable levels. In particular, $r\sim10^{-3}$ is reached for $\xi\sim3.29$ (LiteBIRD target), while the Planck bound $r<0.036$ corresponds to $\xi\sim3.62$. Figure~\ref{fig:parameter_space} shows the GW spectrum $h^2\Omega_{\rm GW}$ as a function of frequency $f$ for $\xi=3.0$, $3.29$, and $3.62$ (solid curves). The spectrum exhibits a peaked shape at $f\sim10^{-3}$ Hz, entering the LISA sensitivity band for $\xi\gtrsim3.3$ (corresponding to the LiteBIRD target $r\sim10^{-3}$ from Fig.~\ref{fig:r_vs_xi}). The horizontal dashed line marks the BBN bound ($h^2\Omega_{\rm GW}<1.8\times10^{-6}$)~\cite{Caprini:2018mtu}, and the shaded regions indicate the sensitivity bands of LISA, BBO, ET, $\mu$Ares, and SKA.

The sourced tensor signal is subject to two main constraints. Backreaction of the produced gauge field requires $\rho_{\mathrm{EM}}\ll V_{\mathrm{inf}}$, which for our benchmark gives $\xi \lesssim 4.3$. A stronger bound comes from non-Gaussianity, limiting $\xi \lesssim 3.6$~\cite{Barnaby:2011qe, Planck:2018}. Thus the viable parameter window for an observable signal is $3.29\lesssim \xi \lesssim 3.62$, consistent with both constraints. Because the amplified gauge field is the actual $\rm U(1)_\trm{B-L}$ gauge boson, its couplings to $\rmbl$ charged matter may induce additional damping or thermalisation, further constraining the parameter space; a detailed study is left for future work. We have also verified that the sourced scalar contribution to $\mathcal{P}_\zeta$ remains subdominant to the inflaton vacuum contribution for $\xi\lesssim3.62$, so the spectral index $n_s$ is unchanged from the standard hybrid inflation prediction.

The GW signal can be further enhanced at smaller scales near the end of inflation or even afterwards. As the Hubble rate decreases and the spectator axion evolves more rapidly, $\xi$ may grow, making gauge field production more efficient. If the axion continues to oscillate after inflation, the Chern–Simons interaction can continue to pump energy into the $\rmbl$ gauge field, thereby enhancing the tensor signal on sub-CMB scales. A detailed frequency-dependent analysis of this post-inflationary contribution is beyond the scope of this work, but we note it as a promising extension.

\begin{figure}[th!]
\centering
\includegraphics[width=\columnwidth]{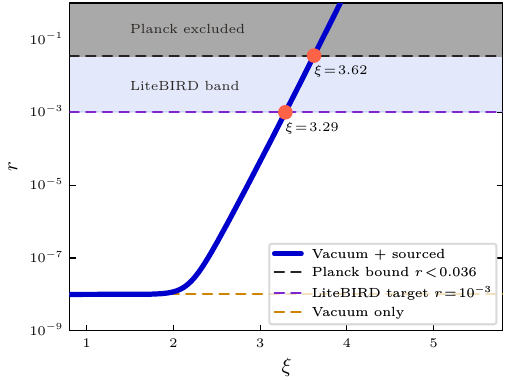}
\caption{Tensor-to-scalar ratio $r$ as a function of $\xi$ (solid curve). Horizontal lines indicate the Planck limit, the LiteBIRD target, and the vacuum prediction.}
\label{fig:r_vs_xi}
\end{figure}

\textit{Discussion and conclusions---}We have shown that supersymmetric $\rmbl$ hybrid inflation can generate an observable gravitational-wave signal despite its negligible vacuum tensor contribution. The mechanism relies on a spectator axion-like field coupled to the $\rm U(1)_\trm{B-L}$ gauge boson through a Chern–Simons interaction. A rolling axion background induces a tachyonic amplification of one helicity of the gauge field, whose anisotropic stress sources tensor perturbations. Our analysis incorporates the relevant theoretical and observational constraints including backreaction and non-Gaussianity, which together delineate the viable parameter window.

From the production mechanism we obtain a sourced tensor spectrum $\mathcal{P}_{\rm T}^{\rm src} \propto e^{4\pi\xi}/\xi^6$, where $\xi = \alpha\dot{\phi}/(2f_a H)$. This exponential dependence allows a significant enhancement even for moderate values $\xi\sim 3.3\text{--}3.6$. The resulting tensor signal may reach $r\sim10^{-3}$, potentially accessible to LiteBIRD, while for $\xi\sim3.6$ the stochastic background enters the LISA sensitivity band, peaking at frequencies around $10^{-3}$ Hz with amplitude $h^2\Omega_{\rm GW}\sim2.5\times10^{-13}$. The mechanism also predicts chiral GW that are highly circularly polarised and a violation of the standard consistency relation $r=16\epsilon$, offering a distinctive observational signature differentiating our setup from standard single-field inflation.

Our framework is a minimal extension of hybrid inflation. It requires no additional hidden sector, leveraging the existing $\rmbl$ gauge field. The hybrid sector drives inflation and background evolution, while the axion–gauge spectator sources the GW signal. Because the axion is a spectator, inflationary dynamics remain unchanged, preserving the successful predictions of $\rmbl$ hybrid inflation for the scalar spectral index and its running. The predicted chirality and the broken consistency relation $r = 16\epsilon$ offer a falsifiable hallmark; a positive detection of both would strongly discriminate this mechanism from standard single‑field inflation.

Looking forward, the interplay between CMB $B$-mode searches and direct GW interferometers provides a powerful multimessenger test. A positive detection in both channels with the correlated amplitude we have discussed would strongly point to an axion–gauge origin. Conversely, null results will push the allowed parameter space to very small $\xi$, reverting to the standard vacuum prediction. Future CMB Stage‑4 experiments and next‑generation interferometers (e.g., Cosmic Explorer, Einstein Telescope) will further extend the reach, potentially probing $\xi$ down to $\sim1.5$ and testing the model over a broader range of scales. Thus, our mechanism offers a testable connection between particle‑physics motivated inflation and future gravitational‑wave observations.
\vskip 0.1cm
\textit{Acknowledgements---}This work is supported by the Science, Technology \& Innovation Funding Authority (STDF) under grant number 48173.


\bibliography{references.bib}
\bibliographystyle{bibi}
\end{document}